\documentclass[aps,pra,twocolumn,groupedaddress]{revtex4}

\usepackage{graphicx}
\usepackage{dcolumn}
\usepackage{bm}
\usepackage{amsmath}

\begin{document} 

\textunderscore

\title{Zero-phonon lines in the spectra of dysprosium atoms in superfluid helium}



\author{P. Moroshkin$^{1,2}$}
\email[]{petr.moroshkin@oist.jp}
\author{K. Kono$^{1,3,4}$}

\affiliation{$^{1}$RIKEN, CEMS, 2-1 Hirosawa, Wako, 351-0198 Saitama, Japan}

\affiliation{$^{2}$ Okinawa Institute of Science and Technology, 1919-1 Tancha, Onna-son, 904-0495 Okinawa, Japan}

\affiliation{$^{3}$International College of Semiconductor Technology, National Chiao Tung University, Hsinchu 300, Taiwan}

\affiliation{$^{4}$Institute of Physics, Kazan Federal University, Kremlyovskaya st. 18, 420008 Kazan, Russia}

\date{\today}

\begin{abstract}
We present an experimental study of a zero-phonon line (ZPL) in the absorption spectrum of Dy atoms solvated in superfluid He. The dopants reside in nanometer-sized spherical cavities known as atomic bubbles. We observe a temperature-dependent broadening of ZPL in the absorption spectrum. The effect is attributed to the scattering of thermal phonons on the atomic bubble that leads to the dephasing of the Dy transition dipole. The extrapolated ZPL intrinsic spectral width at zero temperature is 2300 times larger than the natural linewidth in a free atom. This can be assigned to a fast radiationless quenching of the upper state of the studied transition.
\end{abstract}


\maketitle


\section{\label{sec:Introduction}Introduction}

Impurity atoms and ions embedded in a solid or a liquid can modify the phonon spectrum of the host substance.
In some cases, they lead to the appearance of so-called pseudolocal phonon modes, \textit{i.e.} phonon wavepackets strongly localized around the impurity center.
Liquid and solid He which is a quantum fluid/solid supports a particular type of pseudolocal modes associated with a peculiar structure of the impurity defects known as atomic bubbles (for a review see \cite{TabbertJLTP1997,MoroshkinPR2008}).
These bubbles have a typical diameter of $\approx$1 nm and are formed around neutral impurity atoms, such as alkali and alkali-earth metals due to the strong repulsion between He atoms and the electronic shells of the impurity.
Similar structures are produced by free electrons (electron bubbles) \cite{CelliPR1968} and by some molecular dopants, such as He$_{2}^{\ast}$ excimer \cite{BenderskiiJCP2002}.
The hydrodynamic model of the atomic bubble leads to the eigenmodes of the bubble interface described by the spherical harmonics $Y_{L,m}(\theta,\varphi)$.
At the same time, these vibrations can be represented as localized phonon wavepackets or pseudolocal modes since their eigenfrequencies overlap with the phonon spectrum of liquid and solid He \cite{MoroshkinEPL2011}.

Besides the direct time-resolved measurements \cite{BenderskiiJCP2002} and numeric calculations \cite{ElorantaCPL2004,ElorantaCP2007}, the oscillations of the atomic bubbles can be investigated spectroscopically, by observing the phonon structure in the absorption and emission spectra of the impurities \cite{MoroshkinEPL2011,MoroshkinPRB2018}.
The electronic transitions of the impurity atom that are accompanied by the excitation of the bubble vibrations result in the formation of phonon wings (PW) in the impurity spectra.
The transitions that excite no phonons contribute to a zero-phonon line (ZPL).
The latter is much narrower than PW.
It can be broadened only by the processes that leave the number of phonons unchanged, e.g. by scattering of the phonons already existing in the matrix.

Well-resolved ZPL and PW structures have been obtained also in the spectra of molecular dopants in superfluid He nanodroplets (for a review see \cite{ToenniesACIE2004,CallegariErnstBook2011}).
However, molecular species possess additional degrees of freedom and a strongly anisotropic interaction with surrounding He atoms, which result in a complicated spectra.
The molecular dopants in the droplets either reside at the surface \cite{HigginsJPCA1998} or produce a non-spherical trapping site inside the droplet that may result in a splitting of the ZPL and the appearance of new spectroscopic features \cite{LindingerPCCP2001,HartmannPCCP2002}.
An additional spectroscopic structure is produced by the molecular rotation \cite{PoertnerJCP2002}.
The size distribution of the droplets leads to the inhomogeneous broadening of ZPL \cite{SlenczkaJCP2001}.
Another limitation of the experiments on He droplets is due to the evaporative cooling of the droplet which leads to the droplet temperature $T$ = 0.37 K.
It is thus impossible to vary the helium temperature and pressure and observe their effect on the impurity spectra.

Spherical atomic bubbles in bulk superfluid He can be described in a frame of a relatively simple hydrodynamic model and are suitable for systematic studies of the bubble-phonon interaction.
Zero-phonon lines in bulk liquid and solid He could only be observed in the spectra of inner-shell transitions of the impurity atoms \cite{IshikawaPRB1997,HuiJLTP2000,MoroshkinPRA2011,MoroshkinJCP2013,MoroshkinPRB2018}.
Transitions of valence electrons typically induce a large displacement of the bubble interface and generate a classical wavepacket of a large number of phonons.
As discussed in \cite{MoroshkinEPL2011}, the corresponding spectra representing a multiphonon PW are strongly broadened and shifted and have no ZPL.
The existing studies of the inner-shell transitions could not resolve the intrinsic spectral width of ZPL due to the power broadening \cite{HuiJLTP2000,MoroshkinPRB2018} and the insufficient spectroscopic resolution \cite{IshikawaPRB1997,MoroshkinPRA2011,MoroshkinJCP2013}.

Recently, we have presented an experimental study \cite{MoroshkinPRB2018} of the spectra of Dy atoms in bulk superfluid $^{4}$He, in particular the profile of the phonon wing associated with the $4f^{10}6s^{2}$ $^{5}I_{8}$ - $4f^{9}5d6s^{2}$ $^{5}K_{7}$ inner-shell transition.
Our results suggest that the spectrum of elementary excitations in the vicinity of the atomic bubble is modified with respect to that in pure bulk superfluid He.
Here we present an extension of that study with a special emphasis on the zero-phonon line corresponding to the same electronic transition.
We resolve the intrinsic ZPL spectral width and study its dependence on the liquid He temperature.
The paper is organized as follows: in Sec. \ref{sec:Experiment} we describe our experimental setup and the measurements.
In Sec \ref{sec:Discussion} we discuss our results and compare them to the predictions of the atomic bubble model.
Sec. \ref{sec:Conclusion} gives a summary and conclusions.

\section{\label{sec:Experiment}Experiment}

\subsection{\label{sec:Setup}Experimental setup}

\begin{figure}
	\includegraphics[width=\columnwidth]{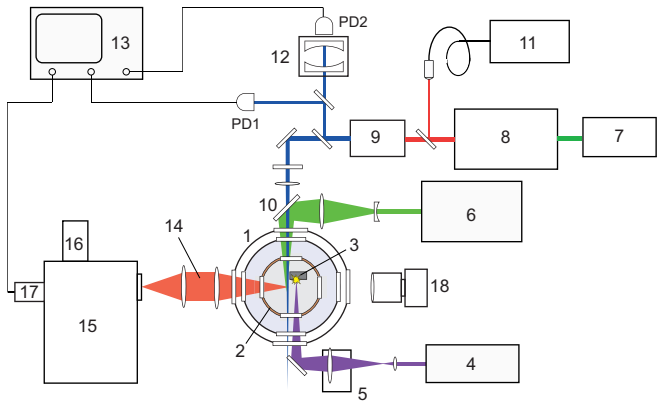}
	\caption{Experimental setup. 1 - cryostat, 2 - sample cell, 3 - ablation target, 4 - pulsed DPSS laser ($\lambda$ = 355 nm), 5 - motorized XY translation stage, 6 - frequency-doubled pulsed Nd:YAG laser ($\lambda$ = 532 nm), or frequency-tripled Nd:YAG laser ($\lambda$ = 355 nm), 7 - cw DPSS laser ($\lambda$ = 532 nm), 8 - cw tunable Ti:Sapphire laser, 9 - second harmonic generator (SHG), 10 - dichroic mirror, 11- wavelength meter, 12 - scanning confocal Fabry-Perot etalon, 13 - oscilloscope, 14 - laser-induced fluorescence, 15 - grating spectrograph, 16 - CCD camera, 17 - PMT, 18 - video camera, PD1 and PD2 - photodiodes.} \label{fig:Setup}
\end{figure}

The experimental setup is described in our recent publication \cite{MoroshkinPRB2018}.
The experiments are carried out in an optical helium-bath cryostat cooled to 1.35--2.1 K by pumping on the helium bath.
The top view of the cryostat with the sample cell and the optical setup is shown in Fig. \ref{fig:Setup}.

Superfluid He in the sample cell is doped with dysprosium atoms by means of laser ablation using two nanosecond pulsed lasers.
The primary ablation of a metallic Dy target (3 in Fig. \ref{fig:Setup}) by a frequency-tripled DPSS laser ($\lambda = 355$ nm) produces mostly metal clusters and nanoparticles.
Dy atoms are produced by the secondary ablation/sputtering of these nanoparticles by another, more powerful pulsed laser (6 in Fig. \ref{fig:Setup}). 
The primary ablation laser (4 in Fig. \ref{fig:Setup}) has a repetition rate of 20--50 Hz and a pulse energy of 70 $\mu$J.
It is focused on the target by a $f$ = 15 cm lens mounted on a motorized XY translation stage (5 in Fig. \ref{fig:Setup}) that is moving in a plane orthogonal to the laser beam.
In this way we move the ablation spot along the target surface thus avoiding drilling of a crater.
For the secondary sputtering we use either a frequency-doubled Nd:YAG laser ($\lambda = 532$ nm) with a repetition rate of 10 Hz and a pulse energy of 0.5--15 mJ, or a frequency-tripled Nd:YAG laser ($\lambda = 355$ nm) with a repetition rate of 20 Hz and a pulse energy of 1.5--6.5 mJ.
It is focused in the middle of the sample cell, above the Dy target.
The ablation process is monitored with a fast digital video camera (18 in Fig. \ref{fig:Setup}) oriented orthogonal to the laser beams and operated at a frame rate of 500--8500 fps.

Dy atoms in liquid He are excited by a second harmonic of a tunable cw Ti:Sapphire laser (8 in Fig. \ref{fig:Setup}) superimposed on the secondary sputtering laser beam using a dichroic mirror.
The laser is tuned into the resonance with the transition from the $4f^{10}6s^{2}$ $^{5}I_{8}$ ground state of Dy towards the state $4f^{9}5d6s^{2}$ $^{5}K_{7}$ at $\lambda=458.9$ nm.
The fundamental wavelength of the Ti:Sapphire laser is measured by a wavelength meter (11 in Fig. \ref{fig:Setup}) with an absolute accuracy of $\pm3.5\times10^{-4}$ nm (0.5 GHz).
The second harmonic linewidth measured by a Fabry-Perot etalon (12 in Fig. \ref{fig:Setup}) does not exceed 300 MHz.
Laser-induced fluorescence is collected at a right angle with respect to the laser beams and is analyzed with a grating spectrograph (15 in Fig. \ref{fig:Setup}) equipped with a CCD camera and a photomultiplier tube (PMT).

\subsection{\label{sec:Results}Experimental results}

The spectrum of the laser-induced fluorescence has been investigated in details in \cite{MoroshkinPRB2018}.
In total we observe 7 spectral lines originating from the electronic states of Dy lying below the laser-excited $4f^{9}5d6s^{2}$ $^{5}K_{7}$ state.
The emission spectrum is dominated by a strong line at 641 nm ($\lambda_{free}$ = 642.4 nm) which originates from the state $(^{5}I_{8})(^{3}P_{0})$, the lowest in the group of $4f^{10}6s6p$ $(^{5}I_{8})(^{3}P_{J})$ states.

The excitation spectrum was obtained by tuning the wavelength of the Ti:Sapphire laser and recording the fluorescence yield of this strongest emission line.
In the series of measurements reported in \cite{MoroshkinPRB2018} we changed the laser frequency in steps of 3--10 GHz and recorded the fluorescence spectrum at each step using a CCD camera.
In this way we have covered the whole excitation spectrum of the $4f^{10}6s^{2}$ $^{5}I_{8}$ - $4f^{9}5d6s^{2}$ $^{5}K_{7}$ transition that consists of a sharp zero-phonon line and a broader phonon wing.
We have demonstrated \cite{MoroshkinPRB2018} that the phonon wing is blueshifted with respect to ZPL by approximately 170 GHz, with a characteristic gap between ZPL and PW arising due to the peculiar structure of the spectrum of elementary excitations (phonons and rotons) in superfluid He.

\begin{figure}
	\includegraphics[width=\columnwidth]{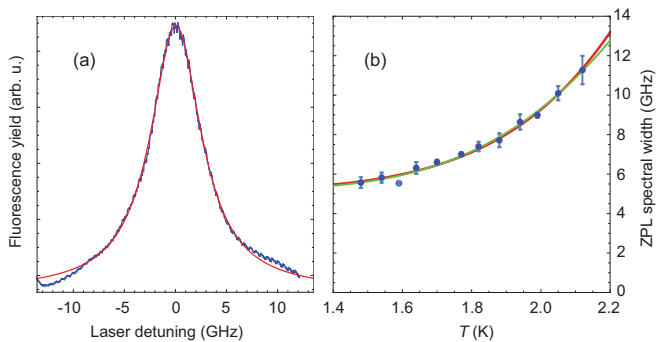}
	\caption{(a) High-resolution scan of ZPL in the excitation spectrum. $T$ = 1.5 K. Dots - experimental data, solid red line - fitted Lorentzian. (b) Temperature dependence of the ZPL spectral width (FWHM). Dots - experimental data, solid red line - fit according to Eq. (\ref{eq:FitPower7}), solid green line - fit according to Eq. (\ref{eq:FitArrhenius}).} \label{fig:ZPLspec}
\end{figure}

In the new series of experiments we concentrate on measuring the lineshape of the zero-phonon line with a higher resolution.
The frequency of the Ti:Sapphire laser was tuned continuously at a rate of 1-2 GHz/s and the time-resolved fluorescence signal at 641 nm was recorded by a photomultiplier tube mounted behind the exit slit of the spectrograph. 
The resulting excitation spectrum was averaged over a large number of frequency sweeps in order to suppress the fluctuations of the fluorescence yield due to the variations of the Dy atomic density.
It was also corrected for the variations of the Ti:Sapphire power during the sweep which was recorded in parallel by a photodiode (PD1 in Fig. \ref{fig:Setup}).
The linearity of the sweep and its amplitude was controlled by recording the fringes of the Fabry-Perot etalon (12 in Fig. \ref{fig:Setup}).

Typical experimental ZPL lineshape is shown in Fig. \ref{fig:ZPLspec}(a).
The spectrum is fitted with a Lorentzian that is shown in the same figure by a solid red line.
The FWHM spectral width $\Delta\nu_{ZPL}$ extracted from the fit lies in the range of 5--20 GHz and increases with the liquid helium temperature as is shown in Fig. \ref{fig:ZPLspec}(b).
Increasing the excitation laser power leads to the saturation of the atomic absorption line and to the spectral broadening of ZPL.
The data reported in Fig. \ref{fig:ZPLspec} have been obtained in the limit of the low excitation power.

\section{\label{sec:Discussion}Discussion}

\subsection{\label{sec:ZPLBroadening} Spectral broadening of ZPL}

The electronic transitions contributing to the zero-phonon line occur without the vibrational excitation of the atomic bubble.
As a result, no phonons or rotons are excited.
The observed temperature-dependent broadening of ZPL in the excitation spectrum can be attributed to the dephasing of the transition dipole of the Dy atom due to the elastic scattering of phonons and rotons already existing in the liquid.

The theory of the ZPL broadening by the scattering of phonons had been developed in \cite{McCumberJAP1963,SmallCPL1978,HsuJCP1984a,HsuJCP1985} for impurity atoms and ions in classical crystalline solids.
The theory predicts a Lorentzian lineshape, in agreement with our observations.
The temperature dependence of the ZPL spectral width $\Delta\nu_{ZPL}(T)$ at low temperatures is determined by the type of the phonons producing the dephasing.
For acoustic phonons, with the density of states described by the Debye model, $\Delta\nu_{ZPL} \propto T^{7}$ \cite{HsuJCP1984b,OsadkoPR1991}.
On the other hand, if the dephasing is due to a pseudolocal phonon mode with a frequency $\Omega$, the broadening is described by Arrhenius law: $\Delta\nu_{ZPL} \propto e^{-\hbar \Omega/k_{B} T}$ \cite{HsuJCP1985,OsadkoPR1991}.

It is not clear a priori, which type of the temperature dependence should be expected for the ZPL of Dy in superfluid He.
The dispersion diagram of the elementary excitations is shown in Fig. \ref{fig:SigmaDisp}.
Here, $\omega$ is the excitation frequency and $k$ is the wave vector.
At low temperatures the spectrum is dominated by acoustic phonons corresponding to the linear part of the dispersion curve at low $k$.
Rotons represent another type of excitations corresponding to the part of the dispersion curve near its minimum at $k$ = 1.9 \AA{}$^{-1}$ \cite{DonnellyJPCRD1998}.
The latter have a well-defined frequency $\omega_{r}/2\pi$ = 0.18 THz and therefore are expected to give a contribution similar to that of the local modes: $\Delta\nu_{ZPL} \propto e^{-\hbar \omega_{r}/k_{B} T}$.

As discussed in \cite{MoroshkinPRB2018}, Dy atom in liquid He is surrounded by a spherical bubble-like void which we call an atomic bubble.
The parameters of this bubble have been computed in \cite{MoroshkinPRB2018} in the frame of a standard spherical atomic bubble model \cite{KinoshitaPRA1995,MoroshkinPR2008}.
The bubble corresponding to the electronic ground state has an equilibrium radius $R_{b} = 5.3$ \AA{}.
In the electronically excited state it expands by approximately 0.15 \AA{}.
The computed undamped eigenfrequencies of the breathing and quadrupolar oscillations of the bubble shape are $\Omega_{0}/2\pi =$ 180 GHz and $\Omega_{2}/2\pi =$ 330 GHz, respectively \cite{MoroshkinPRB2018}.
These frequencies lie within the spectrum of the elementary excitations of bulk superfluid He.
The bubble vibrations thus can be represented as wavepackets of phonons localized around the impurity atom which are referred to as pseudolocal modes.

The experimentally measured $\Delta\nu_{ZPL}(T)$ data in Fig. \ref{fig:ZPLspec} have been fitted with both models:
\begin{align}
\Delta\nu_{ZPL}^{(1)} = \Delta\nu_{0}^{(1)} + A \cdot T^{7} + B \cdot e^{-\frac{\hbar \omega_{r}}{k_{B}T}} \label{eq:FitPower7} \\
\Delta\nu_{ZPL}^{(2)} = \Delta\nu_{0}^{(2)} + C \cdot e^{-\frac{\hbar \Omega}{k_{B}T}}  \label{eq:FitArrhenius}
\end{align}
with adjustable parameters $\Delta\nu_{0}$, $A$, $B$, $C$, and $\Omega$.
The fits are shown in Fig. \ref{fig:ZPLspec}(b) by solid lines.
Due to the small temperature range accessible in the experiment, the data can be fitted by both models reasonably well.
The roton contribution in Eq. (\ref{eq:FitPower7}) turns out to be negligibly small and setting $B$ = 0 only improves the uncertainties of $\Delta\nu_{ZPL}^{(1)}$ and $A$.

The fit with Eq. (\ref{eq:FitArrhenius}) returns the value of the frequency of the pseudolocal mode $\Omega/2\pi = 280 \pm 30$ GHz that is significantly larger than the roton frequency and lies in between the frequencies of the breathing and quadrupolar vibrations.

At $T$ = 0 both models extrapolate to $\Delta\nu_{0} \approx$ 5.1 GHz.
This value is significantly larger than the experimental resolution determined by the laser linewidth and therefore represents the intrinsic linewidth of the transition.
Note that the natural linewidth $\Delta \nu_{nat}$ of the $^{5}I_{8} - ^{5}K_{7}$ transition of Dy ($\lambda_{free} = 458.9$ nm) determined by the radiative decay rate of the upper state \cite{NIST_ASD} is $\Delta \nu_{nat}$ = 2.2 MHz, \textit{i.e.} three orders of magnitude smaller than $\Delta\nu_{0}$.
The observed large spectral width can be attributed to the fast quenching of the laser-excited $^{5}K_{7}$ state by radiationless transitions towards the lower-lying excited states, in particular to the $(^{5}I_{8})(^{3}P_{0})$ state which produces the most intense line in the emission spectrum.

\subsection{\label{sec:AtomicBubble} Scattering of phonons by atomic bubbles}

In this section we consider the interaction of the atomic bubble containing a Dy atom with the elementary excitations of superfluid helium.
We use the parameters of the atomic bubble calculated in our earlier publication \cite{MoroshkinPRB2018}.
Our analysis is based on an acoustic model describing the scattering of sound waves on a classical macroscopically large bubble in a liquid \cite{PaoJAP1963}.
In the past, this approach was successfully applied \cite{CelliPR1968,BaymPRL1969} to describe the interaction of phonons with free electron bubbles in liquid He.

The effective cross section of the phonon scattering by the bubble is calculated as a sum over partial waves:
\begin{equation}
\sigma(k,\theta) = \frac{1}{k^{2}} \left| \sum _{L=0}^{\infty} (2L+1) P_{L}(\cos \theta) f_{L}(k) \right|^{2}  \label{eq:CrossSection}
\end{equation}
With a partial wave amplitude:
\begin{equation}
f_{L}(k) = i \frac{j_{L}'(kR_{b}) + G_{L} k \rho_{He} v^{2} j_{L}(kR_{b})}{h_{L}'(kR_{b}) + G_{L} k \rho_{He} v^{2} h_{L}(kR_{b})}  \label{eq:PartialWaveAmplitude}
\end{equation}
Here, $\rho_{He}$ is the liquid He density, $v$ is the speed of sound, $R_{b}$ is the equilibrium bubble radius, $k$ is the phonon wave vector, $\theta$ is a scattering angle, $j_{L}(x)$ and $h_{L}(x)$ are the spherical Bessel and Hankel functions, $P_{L}(x)$ is Legendre polynomial.
We consider $L = 0$ (breathing) and $L = 2$ (quadrupolar) bubble oscillation modes.

Parameters $G_{L}$ describe the elasticity of the bubble with respect to the corresponding deformation mode.
$G_{0}$ and $G_{2}$ are obtained by introducing extra pressure at the bubble interface $\delta p_{L}(\theta) = p_{L}P_{L}(\cos \theta)$ and computing the resulting bubble deformation $R(\theta) = R_{b} + R_{L} P_{L}(\cos \theta)$.
\begin{equation}
G_{L} = \frac{R_{L}}{p_{L}} \label{eq:BubbleElasticity}
\end{equation}
The calculated values $G_{0} = 0.0126$ \AA{}/bar and $G_{2} = 0.0091$ \AA{}/bar are $\approx$100 times smaller than those obtained for a free electron bubble in \cite{BaymPRL1969}.
The resulting scattering cross section for the atomic bubble is close to that of a hard sphere of the same radius.

\begin{figure}
	\includegraphics[width=\columnwidth]{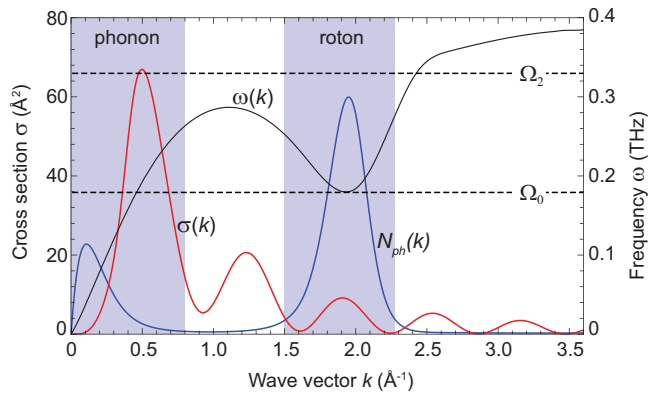}
	\caption{Dispersion diagram of superfluid He $\omega(k)/2\pi$ \cite{DonnellyJPCRD1998} (black, right axis); calculated scattering cross section $\sigma (k)$ (red, left axis) and the density of elementary excitations at $T=1.5$ K (blue, arb. units). Two shaded bands correspond to the two types of elementary excitations: phonons and rotons.} \label{fig:SigmaDisp}
\end{figure}

The cross section including breathing and quadrupolar vibration modes and integrated over the scattering angle is plotted in Fig. \ref{fig:SigmaDisp} as a function of the phonon wave vector.
$\sigma(k)$ has a peak at $k\approx0.5$ \AA{}$^{-1}$ that closely corresponds to the wave vector of the phonons resonant with the bubble breathing vibration.
The density of elementary excitations in the $k$-space at the absolute temperature $T$ is given by
\begin{equation}
N_{ph}(k) = 4 \pi k^{2} \left(\exp \left[ \frac{\hbar\omega(k)}{k_{B}T}\right] - 1 \right)^{-1} \label{eq:PhononDensity}
\end{equation}
In Fig. \ref{fig:SigmaDisp} $N_{ph}(k)$ is shown on an arbitrary scale for $T$=1.5 K. 
It has two maxima corresponding to the phonon and roton branches of the dispersion diagram.
The rate of quasiparticle scattering by the bubble is
\begin{equation}
\Gamma(T) = \int \sigma(k) v_{g} N_{ph}(T,k) dk \label{eq:ScatteringRate},
\end{equation}
where $k_{B}$ is the Boltzmann constant and $v_{g}=d\omega/dk$ is the group velocity.

The dephasing of the atomic transition dipole by the uncorrelated scattering events leads to a Lorentzian lineshape with a FWHM spectral width equal to $\Gamma/\pi$.
The effect is analogous to the so-called impact broadening mechanism \cite{AllardRMP1982} in the gas phase, where the elastic collisions between the atoms lead to the dephasing and to the spectral line broadening.
Here, it is assumed that each scattering event leads to a sudden change of the transition phase by more than 1 radian \cite{AllardRMP1982}. 

\begin{figure}
	\includegraphics[width=0.6\columnwidth]{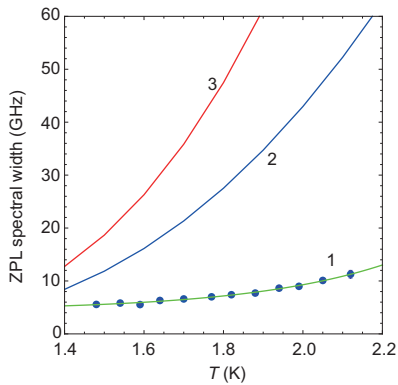}
	\caption{Temperature dependence of the ZPL spectral width (FWHM). Dots and solid line 1 (green) - experimental data, curve 2 (blue) - calculation according to the atomic bubble model, Eqs. (\ref{eq:CrossSection}) - (\ref{eq:ScatteringRate}) including only the excitations from the phonon branch, curve 3 (red) - calculations including both phonons and rotons.} \label{fig:ZPLWidthTheor}
\end{figure}

In Fig. \ref{fig:ZPLWidthTheor} we compare the calculated ZPL spectral width with the experimental data of Fig. \ref{fig:ZPLspec}(b).
Curve 2 is computed by taking into account only the excitations from the phonon branch, $k<1.0$ \AA{}$^{-1}$.
Curve 3 includes both phonons and rotons.
Both calculated dependencies significantly overestimate the experimental data.
At a low temperature the calculated linewidth exceeds the measured value approximately by a factor of two.
As the temperature is increased, the calculated scattering rate increases significantly faster than the experimental line width and the discrepancy increases.
This discrepancy suggests that only a small fraction of thermal phonons and rotons scattered by the bubble leads to a dephasing of the transition dipole.

Scattering of thermal phonons is also responsible for the depolarization of impurity spins and for the broadening of impurity magnetic resonance spectra in liquid and solid He \cite{KinoshitaPRB1994,ArndtPRL1995,FurukawaPRL2006,MoroshkinPR2008}.
In that case, the coupling is very much weaker leading to the spectral line widths of order of 1 Hz.
A more detailed microscopic model of the impurity-phonon (roton) interaction is required for a quantitative interpretation of optical dephasing and spin depolarization data.

\section{\label{sec:Conclusion}Conclusions}

We have investigated the absorption spectrum of the $4f^{10}6s^{2}$ $^{5}I_{8}$ - $4f^{9}5d6s^{2}$ $^{5}K_{7}$ inner-shell transition of Dy atoms embedded in superfluid $^{4}$He.
We have measured for the first time the intrinsic spectral width of the zero-phonon line of an atomic impurity in liquid He and have studied its dependence on the helium temperature in the range of 1.35--2.1 K. 
The observed temperature-dependent broadening of ZPL is attributed to the dephasing of the atomic transition dipole by the scattering of thermal phonons on the impurity atom.
The experimental data do not allow one to determine whether the dephasing is caused by acoustic phonons or by a pseudolocal mode corresponding to the vibrations of the atomic bubble.
However, the effect of rotons seems to be negligible. 
Intrinsic spectral width of ZPL obtained by the extrapolation to $T$ = 0 is three orders of magnitude larger than the natural linewidth in a free atom.
It is attributed to the shortening of the excited state lifetime by a radiationless quenching.

\begin{acknowledgments}
	
This work was supported by JSPS KAKENHI grants No JP24000007 and JP17H01145.
	
\end{acknowledgments}


\end{document}